\documentclass{iopart}
\usepackage{graphicx}
\usepackage{amssymb}
\begin{document}
\title{Strangelets: Who is Looking, and How?}
\author{Evan Finch}
\address{Department of Physics, Yale University, New Haven, CT 06520}
\ead{evan.finch@yale.edu}

\begin{abstract}

It has been over 30 years since the first suggestion that the true ground state of
cold hadronic matter might be not nuclear matter but rather strange
quark matter (SQM).  Ever since, searches for stable SQM have been proceeding
in various forms and have observed a handful of interesting events but
have neither been able to find compelling evidence for stable strangelets nor
to rule out their existence.   I will survey the current status and near future
of such searches with particular emphasis on the idea of SQM from strange star collisions
as part of the cosmic ray flux. 
\end{abstract}

Strange Quark Matter (SQM) is a proposed state of hadronic matter made up
of roughly one-third each of up, down, and strange quarks in a single
hadronic bag that can be as small as baryon number $A=2$ or 
as large as a star.  It was
suggested some 30 years ago that SQM 
(of which a small chunk is called a ``strangelet'')
might in fact be the true ground
state of hadronic matter\cite{bodmer,Witten}.  Whether this is true 
is still an open question today.

The idea that Quark Matter made of only up and down quarks is stable can
be dismissed immediately by the observation that normal 
nuclear matter doesn't decay
into it.  However, in the case of SQM such a decay would require several
simultaneous weak interactions, making it prohibitively unlikely.  
The stability of SQM cannot yet be determined from first principles within 
QCD, but has been addressed in various phenomenological models.  The most commonly used of
these is the MIT Bag Model \cite{chodos,fahri} which also has been extended   
to include the effects of colour superconducting states \cite{alford,MadsenCFL}.  
The results of such calculations are inconclusive, but for a large
part of the ``reasonable'' parameter space in these models, SQM is in fact
absolutely stable for baryon number greater than some minimum value 
(smaller strangelets are disfavored due to curvature energy).
This minimum, depending on parameter choices, is
generally larger than 50 and smaller than 1000 although shell
effects which are important for $A \lesssim 100$ 
may cause islands of stability at smaller $A$ values.  
The key point is that SQM stability is a question that must 
be settled experimentally or observationally.

If it turns out that SQM is stable, the implications would be 
potentially tremendous not only for 
the resultant direct and indirect understanding 
of the strong interaction but also for practical 
applications ranging from new materials (effectively, nuclear charges up to 
$Z\approx 1000$ would become possible) to potential as a clean 
energy source\cite{dalitz}.

\section{Potential sources of Stable SQM}
\subsection{Relics of the Early Universe}
It has been suggested that bubbles of SQM might have formed as the early universe
underwent a first order phase transition from the Quark Gluon Plasma\cite{Witten}, 
and even that the resulting SQM might be the chief component of Dark Matter.
It has since been argued that SQM formed in this manner would not have survived
the conditions of the early universe\cite{Alcock}, but the strongest arguments against
this idea come from experiment as discussed below.

\subsection{Strange Star Collisions}
If Strange Quark Matter is stable at zero pressure, then it is very likely that all
compact stars which are commonly thought of as neutron stars are in fact just big
lumps of SQM called ``strange stars''.  The transition of the
star from normal hadronic matter to strange matter could happen during or after 
the creation of the compact star by various means, but it is 
generally agreed that it likely
would happen if strangelets are stable\cite{madsenSQM}.  This has two important implications.

Firstly, if any compact star can be unambiguously observed to be not a strange star (there are
quite different expectations for the mass-radius relationship of strange stars versus
neutron stars, for example), then we can infer that SQM is likely not absolutely stable.  These
measurements are difficult but are improving \cite{stairs}.  

Secondly, strange stars which exist in binary systems should eventually have collisions 
with their binary partners, likely ejecting some fraction of their mass as strangelets.  
This would lead to a significant flux of strangelets in cosmic rays.  A calculation of the 
resulting flux at Earth  
has been performed \cite{MadsenCR} and leads to the prediction shown as the 
thick dashed line in Figure \ref{fig:madplot}.
\footnote{A calculation of strangelet propagation to Earth has also been done in \cite{Medina-Tanco}.  While no
flux prediction is given, the conclusions are broadly consistent with those in \cite{MadsenCR}}
This prediction has significant uncertainty due to input parameters which are only poorly
known (for example the number of compact stars in binary systems and the average amount
of mass ejected by a strange star collision) and is conservative in most assumptions (for 
example, strangelets are assumed to always be destroyed by nuclear interactions on the
journey to Earth).  The curve represents the expected flux (not including the effect
of the Earth's magnetic field) if all the ejected mass ends
up in the form of strangelets of a single baryon number and in that sense should be considered an 
upper limit.  There is little guidance for the expected mass distribution, although there are some
encouraging indications
that it may peak around $A \approx 1000$\cite{MadsenCFL,KrishnaNew}.  Finally, note that the curve 
shown in Figure \ref{fig:madplot}
assumes the mass-charge relationship $Z = 0.3A^{2/3}$; the flux prediction for  
a given $A$ changes roughly as $Z^{-1.2}$ meaning that it can vary by a factor of a few over a 
reasonable range of values of $Z$.  

The expected strangelet population from this mechanism is less by many ($\approx 12$) 
orders of magnitude than
the necessary population to explain Dark Matter by SQM and so, as discussed in some detail below,
this flux has not
been significantly addressed by any past experiment.  Much of the
rest of this article will focus on the possibility of searching for strangelets at this level.  
Such a search has a real chance for a discovery.  Alternatively, if this expected flux could
be ruled out at a significant level, the hope that SQM is completely stable would be
tremendously reduced. 

\section{Past Experimental Searches and Lessons Learned}
\subsection{Experimental Signatures}
Two key properties of strangelets are exploited for the majority of experimental searches 
\begin {enumerate}
\item With almost equal numbers of up, down, and strange quarks, strangelets would 
have a much lower charge to mass ratio than normal nuclei.
(For the MIT Bag Model $Z \approx 0.1A$ for $A<<10^3$ and 
$Z \approx 8A^{1/3}$ for $A>>10^3$.  With the bag model extended to include colour 
superconductivity called colour-flavour-locking (CFL), $Z \approx 0.3A^{2/3}$.) 
This means that a strangelet of a given
velocity will have a much higher magnetic rigidity ($R=p/Z$) than a normal nucleus of that same 
velocity and also that strangelets will have significantly longer ranges when passing through
material than normal nuclei of similar mass and energy.
\item Strangelets are potentially much more massive than normal nuclei.  This is
exploited for example by looking for Rutherford backscattering of very heavy nuclei
from terrestrial samples and also simply by looking for very dense nuclearites that penetrate
to ground-based or underground detectors\cite{Lowder}.
\end{enumerate}

\subsection{Results relevant to SQM as Early Universe relics.}
SQM nuggets formed as early universe relics would be expected to
have velocities of a few hundred km/sec relative to the Earth
and could be identified by tracks seen in cosmic ray detectors based on the Earth, balloons,
or satellites.  With these velocities the SQM nuclearites would likely be neutral
but would (if sufficiently massive) lose large amounts of energy by elastic scattering
of atoms and molecules in their path, and so leave distinct tracks in cosmic ray detectors.
Also, these nuggets of SQM would be present in the material from which the solar
system was formed and so Atomic Mass Spectroscopy searches for very heavy isotopes of
normal matter are sensitive to this mechanism of SQM formation.

A variety of past experiments have been sensitive to such SQM and these are well summarized in
other reviews \cite{Lowder,Klingenberg}.  The result is that the concentration of SQM 
needed to explain
dark matter as SQM nuggets \cite{glashow} of some mean baryon number $A$ 
is ruled out by 5 or more orders of magnitude
up to approximately $A=10^{20}$ .  

One recent result which is very relevant to this hypothesis is the limit set by searching
seismic records for events consistent with an epilinear source.  A possible candidate
for a SQM nugget weighing approximately a ton (which would be about the size of a red 
blood cell!) had been identified but has now been explained as likely being due to a timing
error in one of the recording stations \cite{herrin}.  The resulting limit becomes the most restrictive
limit to date in the mass range of $A>10^{30}$. 

\subsection{Results relevant to Strangelets from Strange Star collisions}
Strangelets from strange star collisions should be accelerated by the same mechanism as
``normal'' cosmic rays and therefore are expected to follow a similar 
distribution versus rigidity of $dN/dR \propto R^{-2.2}$ \cite{MadsenCR}.  
The flux which reaches the 
Earth would be modulated by (among other things) the solar wind and 
geomagnetic field, effectively killing the flux for $\beta \lesssim 0.2$
for strangelets having a typical charge-mass relationship.

\begin{figure}
\centering
\includegraphics[bb=41 36 543 523,width=400pt]{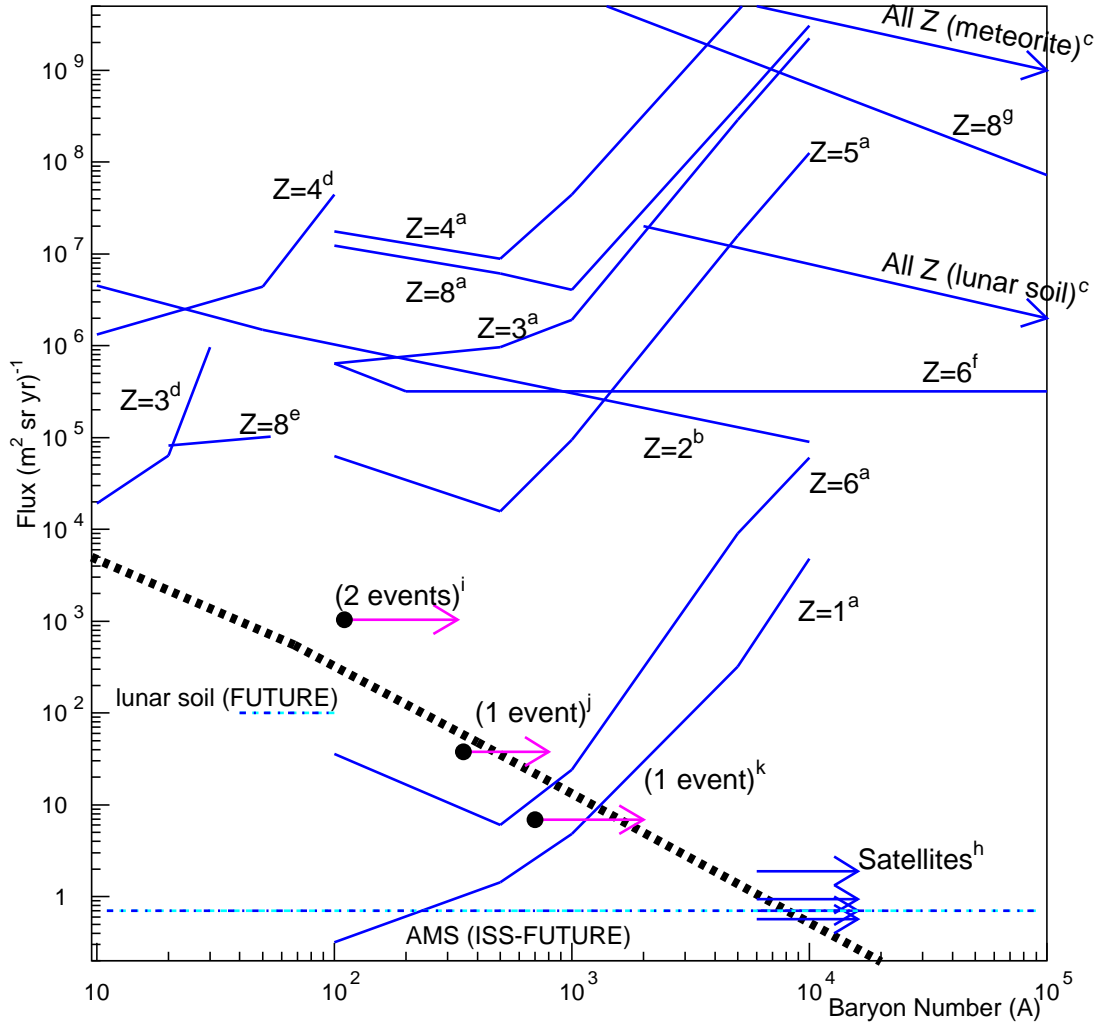}
\caption{\label{fig:madplot}Predicted strangelet flux from strange star collisions 
versus strangelet 
baryon number.  The thick dashed line is the
predicted flux \cite{MadsenCR}.  The solid blue lines
represent previous experimental limits for strangelets which are translated into
flux limits.  
Filled circles represent previous detections
of events consistent with strangelet signals.  The dashed blue lines represent the
expected sensitivity of upcoming searches in lunar soil 
and with AMS. 
Note that there also exists a superior limit for Z=1 which is omitted
as noted in the text.
}
\end{figure}

Experimental limits relevant to the expected flux are shown in Figure \ref{fig:madplot}.
A single figure containing all this information is useful as a general overview, but 
some cautions should be given along with this figure.  First, translating these 
concentration limits into flux limits is very dependent on both the assumed 
strangelet velocity distribution and the flux history; these are
very different depending on whether we are trying to interpret these
results as limits on strangelet flux
from strange stars or population from early universe relics.  Also, as previously mentioned, 
the flux prediction shown on Figure \ref{fig:madplot} only strictly applies to strangelets with the
CFL mass charge relationship. Other sources of uncertainty in these limits are
noted in the following paragraphs.

The curves labeled 'a'\cite{hemmick}, 
'b'\cite{Lu}, 
'd'\cite{Klein}, 
'e'\cite{Middleton}, 
'f'\cite{Turkevich}, 
and 'g'\cite{smith}, 
come from relic searches in terrestrial
material.  To translate these into flux limits, we assume the following:
First, that strangelets are not destroyed as they travel through the atmosphere. 
(Given the expected velocity
distribution discussed above, most of the incident strangelet flux should stop by
ionization energy loss in the atmosphere.)
Second, we assume that strangelets which have been raining down on Earth over its
4 Billion year history have been mixed in the top 20km of the Earth's surface (except for
$Z=2$ strangelets, which we assume behave chemically as Helium and so remain in the atmosphere.)  
Many of these limits apply only to single charge states and are so noted on the figure.  
There have also been crude corrections done to account
for the reduction in flux due to the geomagnetic field; because the low velocity 
flux is already limited by the solar wind, this correction is only as large as a 
factor of 3 for the least rigid charge-mass combinations. 

The curves labeled 'c'\cite{PerilloIsaac} come from searches which bombard
materials with slow moving heavy ions and watch for characteristic gamma rays which
would signal the absorption of the ions by strangelets.  
For lunar soil, we have assumed that strangelets are diluted
by geological mixing only among the top 5 meters of lunar soil while for meteorites 
we have assumed an exposure time of $10^7$ years \cite{monreal}
(compared to $4\cdot10^9$ years for the Earth and moon).

The satellite based searches \cite{ARIEL-6,HEAO-6,Skylab,TREK}(labeled 'h') 
used combinations of Lexan
detectors, Cerenkov counters and scintillators to search for heavily ionizing tracks
above Earth's atmosphere.  The four experiments noted were sensitive to charge
$Z \gtrsim 100$ which for this plot we have converted into 
a lower mass limit following the CFL 
charge-mass relationship noted above.

Finally, three points (labeled 'i'\cite{hecro81},'j'\cite{ETevent}, and 'k'\cite{price}) are noted as 
interesting events.  
These denote four events seen in balloon-borne cosmic ray detectors; each of them 
is consistent with having strangelet characteristics but none provide definitive evidence.  
They are discussed in some detail in \cite{Lowder}.

What then can we conclude from the search limits shown in Figure \ref{fig:madplot}?
We note that the result for charge $Z=1$ strangelets is at a level significantly
below the prediction of flux from strange star collisions.  However, 
very little of the phenomenological
model parameter space which predicts stable strangelets would include stable states 
light enough to have $Z=1$, so this limit is not terribly 
restrictive (in fact, there is an even more 
impressive limit for $Z=1$ \cite{smith} 
which was left off of Figure \ref{fig:madplot} in the interest of readability).  
There is also a $Z=6$ limit which sits just below the flux prediction.  In the mass
range expected for a $Z=6$ strangelet ($40 \lesssim A \lesssim 100$)  
we would expect that there are 
significant shell effects leading perhaps to islands
of stability in $Z$ vs $A$ \cite{madsenSQM} so that this limit, while interesting, 
does not by itself significantly constrain the flux prediction.    
We can conclude that flux at the level of this prediction is still 
essentially an open question.

\section{Upcoming searches}
\subsection{Alpha Magnetic Spectrometer}
The most promising upcoming search relevant to strangelet flux from strange stars
is the Alpha Magnetic Spectrometer (AMS)\cite{AMS} which has been approved to operate aboard
the International Space Station.  AMS will basically be a very high class spectrometer
with an acceptance of $0.4 m^2 sr$
operating above the Earth's atmosphere for measurements of cosmic rays.  For
the strangelet search, the most relevant pieces are
\begin{itemize}
\item A superconducting magnet providing a peak field of .86 T and an analyzing
power of $Bl^2=0.862 Tm^2$. 
\item A silicon strip tracking system consisting of 8 double sided planes
giving position resolution in the bend plane of 8$\mu m$, yielding a rigidity 
resolution of 2.5\% at 100GeV/c.  
Charge can be measured up to $Z \approx 25$.
\item A time of flight system with 4 planes of scintillator measuring time of flight
with an accuracy better than 100ps for $Z>1$ ( $\delta\beta/\beta \approx
0.03$). Charge measurement will be possible up to $Z \approx 20$.  For higher velocity, a 
RICH can be utilized which gives $\delta\beta/\beta \approx 0.001$ .
\end{itemize}
To distinguish a strangelet track from that of a normal nucleus, 
AMS will measure the ratio $R/\beta\gamma (= p/Z \beta \gamma)$.  
This ratio is proportional to $A/Z$ and so 
will be very different for normal nuclei than for strangelets.  AMS should be able to easily
distinguish strangelet events from normal nuclei over a range from $\beta\gamma \approx 0.1$ up
to $R \approx 200 GeV/c$ \cite{Sandweiss}; although
for $Z \gtrsim 20$, AMS will only be able to measure 
the ratio $A/Z$ rather than $A$ and $Z$ individually. 
With the velocity distribution expected of strangelets from strange star 
collisions, AMS is very well suited for this search.

With the expected experiment duration of 3 years on the ISS, AMS will reach the single event 
sensitivity shown as the dashed line in Figure \ref{fig:madplot}.  The sensitivity extends over all
charges and as shown has great potential for discovery if the predicted strangelet
flux exists.  However, given the recent difficulties of the shuttle program, the ultimate
launch of AMS is now somewhat uncertain.

\subsection{Lunar Soil Search using WNSL Tandem}
Because the moon has very little geological mixing compared to the Earth and no magnetic
field, a search for strangelets as a component of lunar soil has an advantage in sensitivity
of about $10^4$ over a similar search with terrestrial material.  Because of this, and motivated
by an interesting event (though again, by no means a definitive strangelet signal) which was 
found in the data taken during the AMS-01 prototype flight \cite{AMS01} aboard the space 
shuttle, we are looking for strangelets in lunar soil using the 
Yale WNSL Tandem Van de Graff accelerator as an atomic mass spectrometer.  In this 
search, the lunar material is ionized (initially, a negative ion is formed) 
and accelerated with the ions undergoing mass and rigidity selection by two dipole magnets.
A stopping foil and final silicon dE-E telescope reduce the remaining background sufficiently
for us to make a meaningful measurement relative to the predicted flux.  Shown on Figure \ref{fig:madplot}
is the sensitivity we hope to achieve in the next year; we currently have achieved a sensitivity
which is worse by about a factor of 20 over a much smaller mass range.  This search will be
sensitive to charge 8 (for which the expected sensitivity is shown on Figure \ref{fig:madplot})
and other nearby charges at varying levels; it is discussed in more 
detail elsewhere in these proceedings\cite{KeHan}.

\subsection{Ground-Based cosmic ray detectors}

The SLIM experiment consists of a large mountaintop-based array of cosmic ray etch detectors.  
The collaboration is in the process of performing a very sensitive search for low velocity strangelets with 
$A > 10^{13}$ (i.e. relevant to the hypothesis of SQM as early universe relics).  As noted in \cite{SLIM},
they will also have a high sensitivity for faster strangelets (i.e. from strange star collisions)
with the important caveat that these strangelets must be many times 
more penetrating than normal nuclei to reach these mountaintop detectors. 
This is not a ridiculous idea since for example it has been suggested that very 
penetrating strangelets might
be an explanation for Centauro events\cite{bjorken}.  With this important caveat then, 
the SLIM collaboration expects an ultimate sensitivity similar to or slightly superior 
to that expected for AMS (for $A$ greater than a few hundred).

A similar experiment is being undertaken using data from the L3 detector with a 
cosmic ray trigger.  This search is discussed elsewhere in these proceedings \cite{XMa}.

\subsection{More Terrestrial searches}

The limit shown in Figure \ref{fig:madplot} for $Z=2$ comes from a 
recent result by a group at Argonne National Lab that 
has done an extremely sensitive search for heavy isotopes of Helium using laser spectroscopy
methods \cite{Lu} which leads to a very strict limit on $Z=2$ strangelet relics of the early
universe over a large mass range. 
When interpreted as a limit on flux from strange star collisions, it does not quite achieve
sensitivity to the expected flux.  However, the authors believe that they can 
improve the sensitivity by several orders of magnitude 
and can also extend this technique to other elements.

As alluded to previously, a difficulty in determining the sensitivity of 
terrestrial searches to
the potential flux from strange stars is determining what such a flux would do upon 
arrival at Earth.  The subsequent fate of strangelets depends
on many non-trivial factors.  A survey of various possible searches in terrestrial samples 
with an eye to what will be most powerful in addressing this flux has been undertaken
in \cite{monreal}.  Among the interesting possibilities noted are searching
for metallic strangelets in the stratosphere (where the author 
believes they will only be diluted by vaporizing
meteorites) and searching for strangelets as heavy isotopes of elements which  
have no normal stable isotopes.  
Of course none of these searches are easy to do well but they do 
show promise of being extremely sensitive.

\section{Accelerator Searches}

Somewhat off the topic of this survey, but still very interesting, are accelerator 
based searches for small metastable strangelets.  
Because coalescence rates of normal nuclei in high energy heavy-ion collisions 
are significantly less than once assumed\cite{E864}, the
hope of seeing strangelets formed by coalescence is dim.  However, other mechanisms
for strangelet formation in these collisions are possible and so searches, particularly
in the baryon-rich forward rapidity regions of ultra-relativistic collisions \cite{STAR},
continue.

\section{Summary}

If stable strangelets exist, it is likely that due to strange star collisions 
there exist strangelets in the cosmic ray flux.  This flux has been predicted with
conservative (though also somewhat uncertain) assumptions to exist at a level that will
be accessible to upcoming searches.  

The most promising of these is the Alpha Magnetic Spectrometer ISS experiment.  The construction
of AMS is nearly complete and we are hopeful that despite the current uncertainty in the  
space program it will be placed aboard the ISS where it will be sensitive to the
predicted flux for baryon number up to several thousand.  If AMS obtains a null
result for strangelets, this will significantly constrain the possibility of stable strangelets
though it will not rule out the idea.
There are also a variety of ongoing and planned searches which will approach this same level of 
sensitivity for cosmic ray strangelets although for a narrower range of parameter space
than AMS can cover.

Additionally, there are continuing searches for low velocity strangelets which may exist 
as relics of the early universe.  It is unlikely that such SQM accounts for
a significant fraction of dark matter, but lower concentrations are possible.

The payoff for finding strangelets is of course very high, and
so the search goes on!

\end{document}